\documentclass[a4paper,11pt]{article}

\usepackage{amssymb}
\usepackage{amsmath}
\usepackage[english]{babel}
\usepackage{graphicx}
\usepackage[ruled,vlined,linesnumbered]{algorithm2e}
\usepackage{float}

\begin{document}
\title{Supercomputer Environment \\ for Recursive Matrix Algorithms}
%\titlerunning{Supercomputer Environment for Recursive Matrix Algorithms}
\author{G.~I.~Malashonok, A.~A.~Sidko \\%\authorrunning{G.~I.~Malashonok*, A.~A.~Sidko*, }
National University of Kyiv-Mohyla Academy, \\
2 Skovorody vul., Kyiv 04070, Ukraine \\
{malaschonok@ukma.edu.ua, a.sidko@ukma.edu.ua }\thanks{Preprint of the paper  Malaschonok, G.I., Sidko, A.A. Supercomputer Environment for Recursive Matrix Algorithms. {\em Program Comput Soft}. 2022, {\bf 48}, 90--101.}}
\date{}
\maketitle
\begin{abstract}
A new runtime environment for the execution of recursive matrix algorithms on a supercomputer with distributed memory is proposed. It is designed both for dense and sparse matrices. The environment ensures decentralized control of the computation process. As an example of a block recursive algorithm, the Cholesky factorization of a symmetric positive definite matrix in the form of a block dichotomous algorithm is described.  The results of experiments with different numbers of cores are presented, demonstrating good scalability of the proposed solution.
\end{abstract}

\setcounter{page}{1}
%%%%%%%%%%%%%%%%%%%%%%%%%%%%%%%%%%%%%%%%%%%%%%%%%%%%%%%
\section{INTRODUCTION}
%%%%%%%%%%%%%%%%%%%%%%%%%%%%%%%%%%%%%%%%%%%%%%%%%%%%%%%
Modern supercomputer systems containing hundreds of thousands of cores face difficulties in the organization of parallel computations (e.g., see \cite{2}). The three main difficulties are the nonuniform hardware workload, accumulation of errors in the process of computations with large matrices, and possible failures of cores during the computation process.
 
A new software infrastructure for organizing distributed computations, which can be called supercomputer runtime environment, should be created. Such an environment should support the development and efficient execution of programs on modern clusters.

It is known that nowadays the most popular runtime environment is Hadoop. It is based on the MapReduce paradigm: a job is split into a number of identical tasks that are executed on processors of the same cluster; then, the results are joined  \cite{23}, \cite{22}.

 This specific class of problems turned out to be fairly popular. However, Hadoop cannot be used for other types of problems.

 Presently, universal systems, such as OpenMP \cite{15}, StarPU \cite{3}, Legion \cite{14}, OmpSs \cite{9}, OCR \cite{10}, HPX \cite{20}, SuperGlue \cite{15}, QUARK \cite{1}, DPLASMA \cite{6} are being developed. Each of them gives an abstract description of available resources and simplifies the development of parallel programs.
 
Recently, a universal Dynamic Task Discovery (DTD) scheme for the PaRSEC runtime environment \cite{7}, \cite{16} has been developed. This environment can support systems with shared and distributed memory. This new paradigm demonstrated better performance compared with the parameterized task scheduling that was used earlier.

Of particular interest is package \cite{17},  designed for synthesizing parallel programs implementing block recursive algorithms. It makes it possible to calculate the transmission of dense blocks of data for distributed memory. In \cite{4} a methodology for designing block recursive algorithms for computer networks of various configurations was presented. In this methodology, independent computational tasks are selected using the tensor product notation.

In this paper, we propose a new runtime environment called DAP for supercomputers with distributed memory. It is designed for solving matrix problems using block recursive algorithms. Like Hadoop, it is a highly specialized runtime environment.

Its main advantage is to provide an efficient computational process and good scalability of programs both for sparse and dense matrices on a cluster with distributed memory. Another advantage is the ability to reorganize the computational process in the event of failure of individual nodes during computations.
The first approach to the creation of such a runtime environment was the dynamic control scheme LLP developed in the laboratory of algebraic computations at Tambov State University in cooperation with the Institute for System Programming, Russian Academy of Sciences. Here one of the cluster nodes was used as the dispatcher  \cite{21}, \cite{25}.

Later, it was elaborated and a scheme with dynamic decentralized control \cite{11}. was developed. However, this scheme did not take into account the recursion depth and it was impossible to switch to another task until the current task was completed.

  The decentralized control scheme was called DAP (Drop-Amine-Pine) \cite{5}. The main concept of this scheme is a drop, which is a fairly complex computational task that can be separated and executed on another processor. The result of the drop computation must be returned to the processor from which it was sent. The process of drop computation assumes that it is unrolled in depth. A computation graph called amine is created. The vertices of this graph are drops of the next nesting level. All amines computed on the current processor are stored in a special structure called pine. In the process of computations, the block recursive function is unrolled in depth unrolled in depth and all its states at each nested level are stored. This allows each processor to easily switch from one subtask to another without waiting for the current subtask to be completed. Moreover, this approach makes it possible to efficiently balance the workload in the case of irregular and sparse data.
  
The paper is organized as follows. In Section 2, we describe the graph of the recursive algorithm and provide examples of block recursive algorithms, such as multiplication of matrices, inversion of triangular matrices, and a dichotomous block recursive algorithm for the Cholesky factorization. In Section 3, we provide a detailed description of the new runtime environment. 
The main objects, fields, threads, and algorithms of the main procedures and functions are described. Section 4 is devoted to a detailed description of the implementation of a block recursive algorithm in the DAP runtime environment using the Cholesky factorization algorithm as an example. In Section 5, we discuss experimental results. The first series of experiments was carried out for investigating the accumulation of computational errors in direct matrix algorithms that use the double precision floating point numbers 
(IEEE standard 754). In the experiments, we calculated the Cholesky factorization. In the next series of experiments, we demonstrated the 
advantages of the new runtime environment, in particular, good scalability.
The experiments were performed for three algorithm: multiplication of
matrices, inversion of triangular matrices, and Cholesky factorization. Special series of experiments were performed for double precision and BigDecimal data.

\section{GRAPH OF A RECURSIVE ALGORITHM}

Consider some examples of block recursive algorithms because the runtime environment described in the following sections is designed for this type of algorithms. One of them will be used in Section 4 for the detailed description of its implementation in the DAP runtime environment.

We consider three block recursive algorithms. Each of them contains a small number of types of recursive computational blocks-matrix multiplication, 
inversion, and factorization. In all these examples, we assume that square matrices in which the number of rows and columns is a power of two. We partition these matrices into four equal blocks. These blocks can be again 
partitioned into equal blocks, and so on. A matrix of any size can be embedded into such a square matrix and the final result can be extracted at the end.

These algorithms are chosen as simple examples. In fact, the majority of matrix algorithms can be written in block recursive form.

\medskip

2.1. \textit{Matrix Multiplication}

There are two matrices A and B. We want to find their product C = AB. We partition these matrices into four equal blocks, and calculate each of the four blocks in the matrix product separately:
$$ A\cdot B = \ \left(\begin{array}{cc}a & \ b \\ c & \ d \end{array}\right) \ \cdot \ \left(\begin{array}{cc}l & \ m \\ n & \ p \end{array}\right) = \left(\begin{array}{cc}w_1 & \ w_2 \\ w_3 & \ w_4 \end{array}\right)   $$
 
$ 1) w_1 =a\cdot l + b\cdot n $ \ \  $ 2) w_2 =a\cdot m + b\cdot p $

$ 3) w_3 =c\cdot l + d\cdot n $ \ \ $ 4) w_4 =c\cdot l + d\cdot n $

Here we have two types of block recursive operations: (1) multiplication of two blocks AB, and (2) multiplication of two blocks and summation with a third block.

   It is important that the operation of block summation cannot form an independent drop because its computational complexity is low -- the number of summation operations equals the number of elements in each block.
   
The graph of this recursive algorithm is shown in Fig. 1. Here are eight drops -- four drops of type (1) and four drops of type (2). The top oval denotes the input function in which the matrices A and B are partitioned into blocks. The bottom oval denotes the output function that assembles the product matrix C from four blocks.

\begin{figure}[H]
\begin{center}
\includegraphics[scale=0.4]{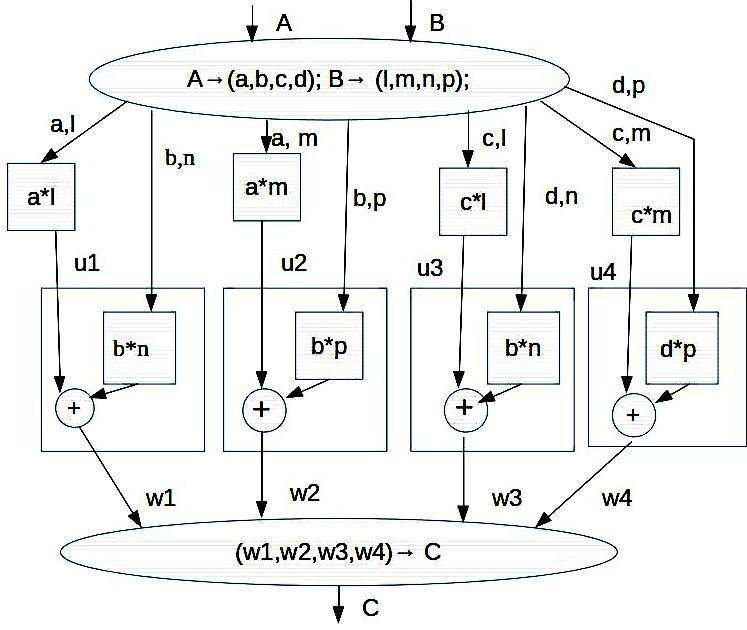}%width=100mm
 %\includegraphics[scale=0.3]{images/upload_file_button}%width=100mm
 % \caption{ Граф блочно-рекурсивного алгорит Mа у Mножения  Mатриц}%

 \label{graphmult}
 \end{center}
\end{figure}

Fig. 1. Graph of the block recursive matrix multiplication algorithm.
%\medskip
	
\subsection{ Inversion of a Triangular Matrix}

A lower triangular matrix A is given that has no zero diagonal elements; therefore, $det(A) \neq 0$. We want to find the inverse matrix $A^{-1}$. 
	
Denote the blocks of the matrix A by $a, 0, c,$ and $d,$; and denote the blocks of the matrix $A^{-1}$ by $x, 0, z, $ and $k.$ The product of these matrices must be equal to the identity matrix:

$ A\cdot A^{-1} = \left(\begin{array}{cc}a & \ 0 \\ c & \ d \end{array}\right) \ \left(\begin{array}{cc}x & \ 0 \\ z & \ k \end{array}\right) \ = \ \left(\begin{array}{cc}I & \ 0 \\ 0 & \ I \end{array}\right) . \ $

\begin{figure}[H]
\begin{center}
\includegraphics[scale=0.4]{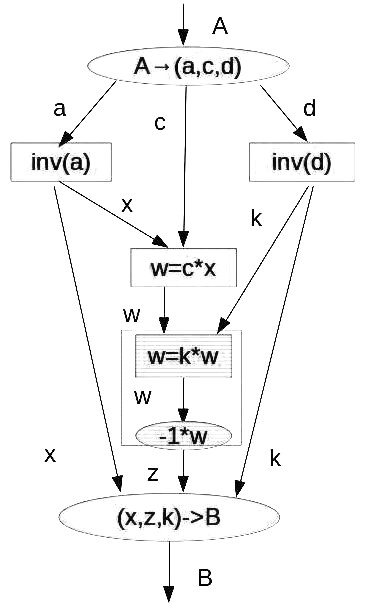}%width=100mm
 %\includegraphics[scale=0.3]{images/upload_file_button}%width=100mm
 % \caption{ Граф блочно-рекурсивного алгорит Mа  обращения треугольной  Mатрицы.}%
 \label{graphinv}
 \end{center}
\end{figure}

Fig. 2. Graph of the block recursive matrix algorithm for inverting a triangular matrix.

\medskip 

Perform the multiplication and equate the corresponding blocks on the left and right sides. This gives the blocks of the inverse matrix:

$ x = a^{-1}, k = d^{-1}, z = -k\cdot c \cdot x $

Here we have three types of block recursive operations: (1) multiplication of two blocks, (2) inversion of a block, and (3) multiplication of two blocks with inversion of the sign of each element. The graph of this recursive algorithm is shown in Fig. 2.

Here we have four drops -- two drops of type (2), one drop of type (1), and one drop of type (3). The top oval denotes the input function in which the matrices A and B are partitioned into blocks. The bottom oval denotes the output function that assembles the inverse from blocks.

	%\medskip

\subsection{ Cholesky Factorization}

  A symmetric positive definite matrix $A$ is given. We want to find a lower triangular matrix $L$ such that 
  $$A = LL^T. \eqno(1) $$  
    In \cite{8} , the standard scheme of Cholesky factorization is described (see Algorithm 1). Note that here the original matrix is partitioned into unequal blocks. The size $n_b \times n_b$ of the block $A_{11}$ in the top left corner must be small such that it can be factorized on one processor: the factorization $A_{11} = L_{11}L_{11}^T $ is performed using a sequential algorithm. Then, the inverse matrix $L_{11}^{-1}$ is calculated by the Gauss algorithm. The main part of the matrix remains in the block $A_{22}$. 
    
%    \subsection{ Стандартный блочный алгорит M разложения Холецкого}

 \begin{algorithm}
\DontPrintSemicolon 
\KwIn{$A$}
\KwOut{ $L$ (see (1))} 
\Begin{
  \For{ each panel left to right}{

\textbf{Partition} $ A = \ \left(\begin{array}{cc}A_{11} & \ A_{21}^T \\ A_{21} &    \ A_{22} \end{array}\right) \ $, \\ where $A_{11}$ is \textbf{$n_b\times n_b$}

\textbf{Factorize} $A_{11} = L_{11}L_{11}^T$ using unblocked algorithm

\textbf{Update} panel $A_{21} = A_{21}L_{11}^{-T}$ using triangular solver

\textbf{Update} trailing matrix  using symmetric rank-k update 
$A_{22}=A_{22}-A_{21}A_{21}^T$

\textbf{Continue} with $A = A_{22}$

 }
  
  \medskip
  }
  
  \caption{Standard block algorithm for Cholesky factorization \cite{8} }
  \end{algorithm}

%\begin{figure}[H] \begin{center}
%\includegraphics[scale=0.46]{images/alg1}%width=100mm
%6 \label{alg1}  \end{center} \end{figure}

We propose a dichotomous block recursive algorithm for the Cholesky factorization. Its feature is that the original matrix is partitioned into four blocks of equal size. In addition, there is no need to invert the matrix 
$L$ using the Gauss algorithm. The computational complexity of the entire
algorithm is the same as that of the matrix multiplication algorithm that 
will be used for block multiplication. Denote the blocks of the matrix by
$ \alpha, \beta, \beta^T, \gamma,$ and denote the blocks of the matrix $L$ to be found by $a, 0, b, c$; and equate the product $LL^T$ to A:
$$L\cdot L^T = \ \left(\begin{array}{cc}a & \ 0 \\ b & \ c \end{array}\right) \cdot \ \left(\begin{array}{cc}a^{T} & \ \ b^{T} \\ \ 0 & \ c^{T} \end{array}\right) \ = $$
$$\left(\begin{array}{cc}aa^{T} & ab^{T} \\ ba^{T} & \ bb^{T} + cc^{T} \end{array}\right) \ =\left(\begin{array}{cc}\alpha & \ \beta \\ \beta^{T} & \ \gamma \end{array}\right)\ = A $$
 This implies the equalities for blocks: 
$$ aa^{T} = \alpha,\ \  b = \beta^{T}(a^{-1})^T, \ \  cc^{T} = \gamma - bb^{T}.$$ 
To calculate the blocks $a$, $b$, and $c$, we must perform block multiplication, transposition, inversion, and Cholesky factorization for two blocks.

The calculation of the inverse matrix can be avoided if the inverse matrix  $L$  is calculated together with the matrix $L^{-1}$. 

 \begin{algorithm}
\DontPrintSemicolon
\KwIn{$A$}
\KwOut{ $\  L,\  L^{-1}$ (see (1))}
\Begin{
        
\If{$size(A) == 1\ \&\ A=[\alpha])$} {
 \KwRet{ $([ \alpha^{1/2}],  [ \alpha^{-1/2}])$ }}   
\Else {

 $A -> (\alpha, \beta, \gamma)$ -- ''we create blocks''
 
 $(a, a_1) = Cholesky(\alpha)$ -- ''$a_1 = a^{-1}$''

 $b^{T} = a_1*\beta; b = (b^{T})^{T}$

 $\delta = \gamma - bb^{T}$

 $(c, c_1) = Cholesky(\delta)$ -- ''$c_1 = c^{-1}$''

 $z = -c_1*b*a_1$

 \KwRet{$ \left(\begin{array}{cc}a & \ 0 \\ b & \ c \end{array}\right) , \ \left(\begin{array}{cc}a_{1} & \ 0 \\ z & \ c_{1} \end{array}\right) $}   
   
 } 
 }
%\medskip
\caption{ Dichotomous  block recursive  algorithm for Cholesky factorization $(L,L^{-1})=Cholesky(A)$}
\end{algorithm}

We extend the procedure for the Cholesky factorization in such a way that it returns not only the matrix $L$ but also its inverse matrix  $L^{-1}$ . Then, together with the blocks a and b, the inverse blocks $a^{-1}$ and $b^{-1}$. In this case, the inverse matrix of
$L$ will have the form:
$$L^{-1} = \ \left(\begin{array}{cc}a^{-1} & \ 0 \\ -c^{-1}ba^{-1} & \ c^{-1} \end{array}\right) \ .$$
             
The graph of this recursive algorithm is shown in Fig.  5.

%\medskip

\subsection{ Cholesky Factorization for the Case $2\times2$ }
 $$\Bigg(\left[\begin{array}{cc}a & \ 0 \\ b & \ c \end{array}\right] , \left[\begin{array}{cc} \frac 1 a & \ 0 \\ \frac{-b}{ac} & \ \frac 1 c \end{array}\right]\Bigg)=Cholesky\left[\begin{array}{cc}\alpha & \ \beta \\ \beta & \ \gamma \end{array}\right] $$
   $$a = \sqrt(\alpha), b = \frac{\beta}{\sqrt(\alpha)}, 
 d = \frac{\alpha*\gamma - \beta^{2}}{\alpha}, c=\sqrt(d).$$

\section{ARCHITECTURE OF THE RUNTIME ENVIRONMENT}
\subsection{Main Steps of the Computational Process. Tree of the Computational Process}
%
%\medskip
%
In all the algorithms described above, the matrix is dichotomously and recursively partitioned into blocks. Each block is associated with a vertex in the graph of the algorithm. The block recursive algorithm is again applied to this block, and this is continued until the blocks become small.

When the block size becomes relatively small, i.e., when the time of data transmission between processors becomes comparable with the computation time, this block becomes a leaf vertex in the algorithm tree, and it will be computed by a sequential algorithm.

The size of a leaf block should be automatically chosen depending on the hardware since it depends on the physical characteristics of computer data transmission rate through the network and the power of processors.

The tree of connections for the computer nodes is formed when the drops are transmitted from the parent to child nodes. The tree of connections is constructed in accordance with the graph of the recursive algorithm and if there are free nodes.

At the initial time, all nodes are free, and a distinguished root node takes on the entire task and the entire list of free nodes.
The computation process consists of three phases.

   \subsubsection{First phase.} 
   In this phase, the tree of connections for the computer modes is initially constructed. The drops with matrix blocks are sent from the root node to child nodes together with the lists of free nodes. The list of free nodes is partitioned into approximately equal parts. From the child nodes, new drops are sent further with the corresponding parts of the list of free nodes. Each time, the set of free nodes is partitioned into approximately equal parts independently of the block size.
Each processor completes the first phase when its list of free nodes is empty or when the amount of data in the received block is less than a certain quantity called "leaf size".

  \subsubsection{Second phase.}
  The processor completes the execution of the first received drop and returns the result to the parent node. If a processor becomes idle either after it sent its result or if the result of its child processor is not yet ready, then this processor adds its index to the list of free nodes and sends it to the parent processor.
  
The list of free nodes may be sent in two directions it may be directed 
either to the parent node or to its child nodes. If some child nodes did not return the result, the list of free nodes may be partitioned into 
approximately equal parts and sent to the child nodes that have drops with the minimal recursion depth, i.e., to the nodes that have blocks of the greatest size. The transmission directions alternate.

A free processor may receive a new drop from any processor that has the index of the former processor in its list of free processors. After completion, the latter processor must return the result to the former processor.

 \subsubsection{  Third phase. } In this phase, the results are returned to the root node, all processors get free, and their indexes are sent to the list of free nodes of the root processor. The result of task execution will be formed at the root node, and the computations will be completed. 

\subsection{Balancing Workload.}

 The runtime environment automatically redistributes subtasks from overloaded nodes to free nodes. For this purpose, a scheme of information transmission about free nodes and overloaded nodes is provided.
 
The information about free nodes is distributed over the lists of free nodes.

The information about the state of child processors is stored in the terminal of child processors and is an integer number showing the depth of the drop recursion that is executed on the child processor.

The child processor keeps this information up-to-date and sends a message when it changes.

The list of free processors is sent to the child processors in accordance with their state: approximately equal parts of the list of free nodes are sent to the child processors that have drops with the lowest recursion level.

\subsection{Components of the Computation Process Control Mechanism}

Consider the components of the computation process control mechanism (Figs. 3 and 4).

\subsubsection{ Drop}

 Drops are parts of the computation graph that are compact subgraphs and can be sent to other processors. The number of machine operations required for the drop execution must be significantly greater than the amount of data at the input and the output. For example, the matrix summation operation cannot be an independent drop.
 
In Figs. 1 and 2, the vertices joined into one drop are included into boxes.

 \begin{figure}[H]
\begin{center}
\includegraphics[scale=0.4]{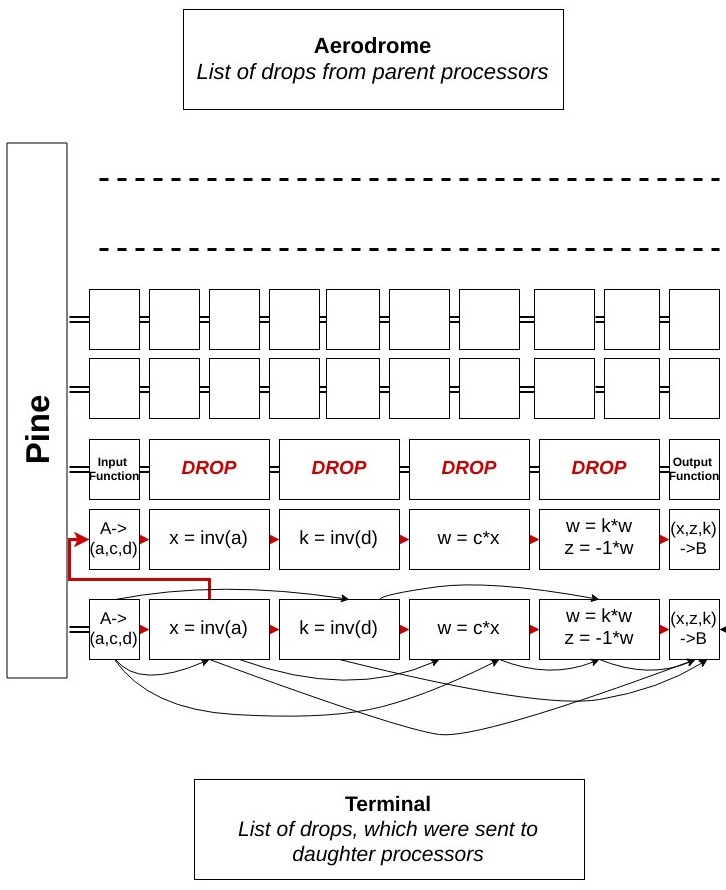}%width=100mm
 %\includegraphics[scale=0.3]{images/upload_file_button}%width=100mm
 % \caption{ Граф блочно-рекурсивного алгорит Mа  обращения треугольной  Mатрицы.}%
 \label{diagr1}
 \end{center}
\end{figure}
Fig. 3. Mechanism for controlling the computational process  Drop, Pine, Amine, Aerodrome, Terminal.

For drops of some types, the vector of input data is divided into two parts  the main and additional ones. For example, the input vector $A*B+C$ for the drop $(A, B, C)$ has the main components $(A, B, -)$ and additional components $(-, -, C)$.  The main components are sufficient for starting the computation of such a drop. The additional components may be delivered later.

Correspondingly, the drop objects can be classified into four types, depending on which data is included in the object: (1) entire input vector, (2) main components of the input vector, (3) additional components of the input vector, (4) result of computations.

\subsubsection{Amine}

 The computational process that executes one drop will be unrolled in depth. During unrolling,  a computation graph called amine is created in which the vertices are the drops of the next nested level.
 
Each amine has one input function, which receives an array of input data and can execute simple preliminary computations. It also has one output function that can execute simple final computations and form the array of output data.

For example, the amine $A * B$ consists of four drops $A * B$, four drops $A * B + C$, one input and output functions.
   \begin{figure}[H]
\begin{center}
\includegraphics[scale=0.5]{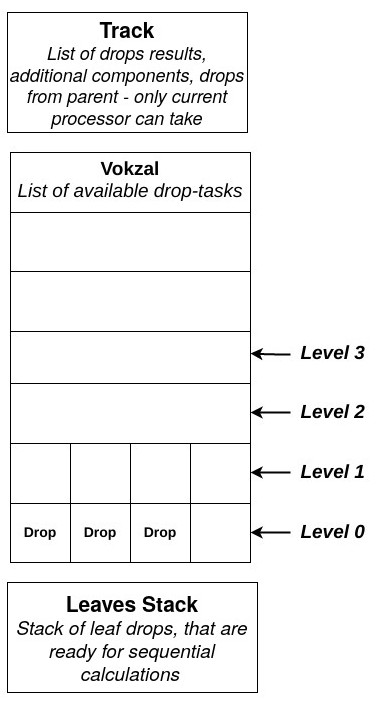}%width=100mm
 %\includegraphics[scale=0.3]{images/upload_file_button}%width=100mm
 % \caption{ Граф блочно-рекурсивного алгорит Mа  обращения треугольной  Mатрицы.}%
 \label{diagr12}
 \end{center}
\end{figure}
Fig. 4. Mechanism for controlling the computational process  Track, Vokzal, LeafStack.
  
\subsubsection{  Pine}

All amines computed on the same processor are stored in a common list called pine (see Fig. 3).

\subsubsection{Vokzal (Station) }

All drops (we will also call them drop tasks) wait in a queue at the vokzal. These drop tasks are at different platforms (or levels), where the level denotes the recursion depth of the drop. A drop can be sent from the vokzal to another processor or it can be sent for execution on the current processor. The vokzal level is the minimum drop level at it.

\subsubsection{Aerodrome}

Each processor that sent a drop to the current processor is its parent processor. The list of all active parent processors is called aerodrome. The parent processor that received the results of all drops sent by it is deleted from the aerodrome.

However, the main parent processor from which the first drop was obtained is always kept since the list of free processors must be sent to this parent processor.

\subsubsection{Terminal}

The terminal is used to communicate with the child processors to which drop tasks were sent. All child processors are registered in the terminal, and their current recursion level is stored in the terminal. The child processors keep it up-to-date and for this purpose they send messages when their recursion level changes.

The child processor that returned the results of all drops tasks received by it is removed from the terminal. The terminal level is the minimum level of the child processors.

\subsubsection{Track}

   This is a platform from which no drop tasks can be sent to other processors.
   
Drops from parent processors are sent to the track, results produced by all drops  those that were executed on this processor and obtained from the child processors  are stored here.

\subsubsection{Leaf stack.}

   Here leaf drop tasks, i.e., those that should be executed on this processors because the amount of data in them is small, are stored.

\subsection{BASIC FIELDS AND FUNCTIONS}

\subsubsection{Fields of the drop object.}
     ---  PAD (np, na, nd) is the drop address, where np is the processor index, na is the amine index, and nd is the drop index. The computation result must be returned to this address.
    
 ---  Type is the type of drop (the unique index in the list of all drop types).

 ---  InData and outData are the vectors of input and output data of the drop. The vector InData has two parts  main and additional components. The main components are sufficient for starting the computations; the additional components may arrive later.

 ---  Amine is the pointer to the amine where the drop resides.

 ---  RecNum is the level index, i.e., the index of the recursion depth.

 ---  Arcs is the graph topology. It is encoded by an integer array; e.g. see the graph topology of the Cholesky factorization algorithm in Fig. 6.

\subsubsection{Fields of the amine object.}

    - PAD (np, na, nd), Type, inData, outData are the same as in the drop object.
    
    - Drop is the array of all drops of this amine.

\subsection{Organization of Two Threads}

Two threads  a computation thread and a dispatcher thread  are used. These threads are executed on each core of the cluster by turns.

\subsubsection{Computation thread.}

This thread waits for the arrival of the first drop task to the station and starts executing.

\medskip

\textit{Objects of the computation thread are as follows:}

 ---  Pine is the list of amines on the processor;

 ---  Vokzal is the array of lists of available drop tasks;

 ---  Aerodrome is the list of parent processors;

 ---  Terminal is the array of lists of child processors;

 ---  CurrentDrop is the currently executing drop.
    
    \medskip

\textit{Functions of the computation thread are as follows:}

 ---  WriteResultsToAmin this function writes the computed output vector of the drop within its amine to the output vectors of other drops according to the graph topology.

 ---  InputDataToAmin creates an amine corresponding to the current drop, calls the input function, and passes the input vector to its entry.

 ---  WriteResultsAfterInpFunc writes the result of the input function computation to all drops of the amine according to the graph topology (see Fig. 5 as an example).

 ---  runCalcThread is the main procedure of the computation thread. While Track is not empty, the results of computing drops are registered according to the topology and the drops ready to be executed are written to the vokzal. If a new drop task arrives, a new amine is created and its input function is computed. While LeafStack is not empty, then the leaf drops are successively executed. While the vokzal is not empty, the drop tasks residing at the vokzal are executed. If all these fields are empty, then the corresponding flag is set and the computation thread goes to the wait state, the index of its processor is added  to the list of free processors, and this list  sends.

\subsubsection{Dispatcher thread.}
The work of the dispatcher thread can be partitioned into ten processes:

1. Waiting for a signal indicating that all computations have been to completed.

2. Receiving a drop task.

3. Receiving the list of free processors.

4. Receiving and writing the state level of the child processor.

5. Receiving the computation result from the child drop writing it into the corresponding amine.

6. Receiving additional components from the child processor.

7. Sending drop tasks from the vokzal to free processors.

8. Sending free processors to child processors or to the parent processor; these actions alternate.

9. Sending the results produced by drop tasks to the parent processors.

10. Sending additional components to the child processors.

The source code of the program is available at the URL \\
{https://bitbucket.org/mathpar/dap/src/master/src/main/java/com/ \\
mathpar/parallel/dap/}

\section{CHOLESKY FACTORIZATION}

   As an illustration, we consider the computation process for the Cholesky factorization algorithm in more detail. 
   
   Figures 5 and 6 show the recursive graph of the algorithm and the corresponding array of arcs representing the graph topology. 
   
All drops in the graph are numbered by natural numbers. Each row in the array Arcs corresponds to one drop. The row index is the drop number. 

Each triple of numbers in a row indicates the relation of the drop with the inputs of the next drops. The first number is the drop number that receives data, the second number is the number of component in the output vector of this drop, and the third number is the number of component in the input vector.

The array Arcs begins with the zero row, which corresponds to the drop input function, and the last row is intended for the output function.

For example, the second row in Arcs is {7 0 0, 2 1 0, 5 1 1, 7 1 3}. Consider the first triple 7 0 0. The number 7 corresponds to the index of the dependent drop into which the result should be written. The two zeros show that the zero component of the output vector is connected to the zero component of the input vector.

The next triple is 2 1 0. The number 2 corresponds to the index of the dependent drop, 1 is the number of component in the output vector, and 0 is the number of component in the input vector. 

      \begin{figure}[H]
\begin{center}
\includegraphics[scale=0.9]{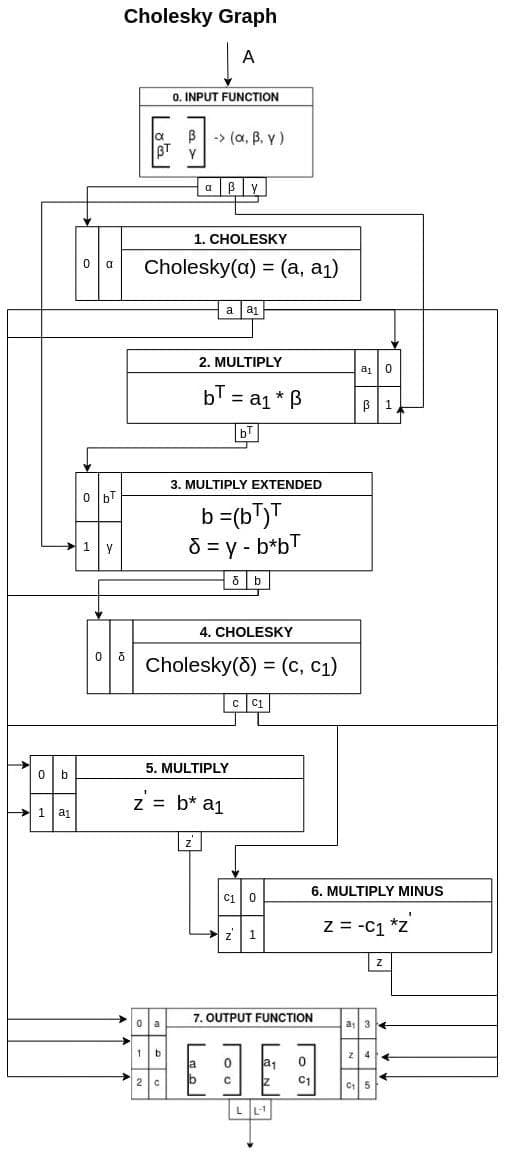}%width=100mm
 %\includegraphics[scale=0.3]{images/upload_file_button}%width=100mm
 % \caption{ Граф блочно-рекурсивного алгорит Mа  обращения треугольной  Mатрицы.}%
 \label{arcs_2}
 \end{center}
\end{figure}

Fig. 5. Graph of Cholesky factorization. 

  \medskip

   \begin{figure}[H]
\begin{center}
\includegraphics[scale=0.20]{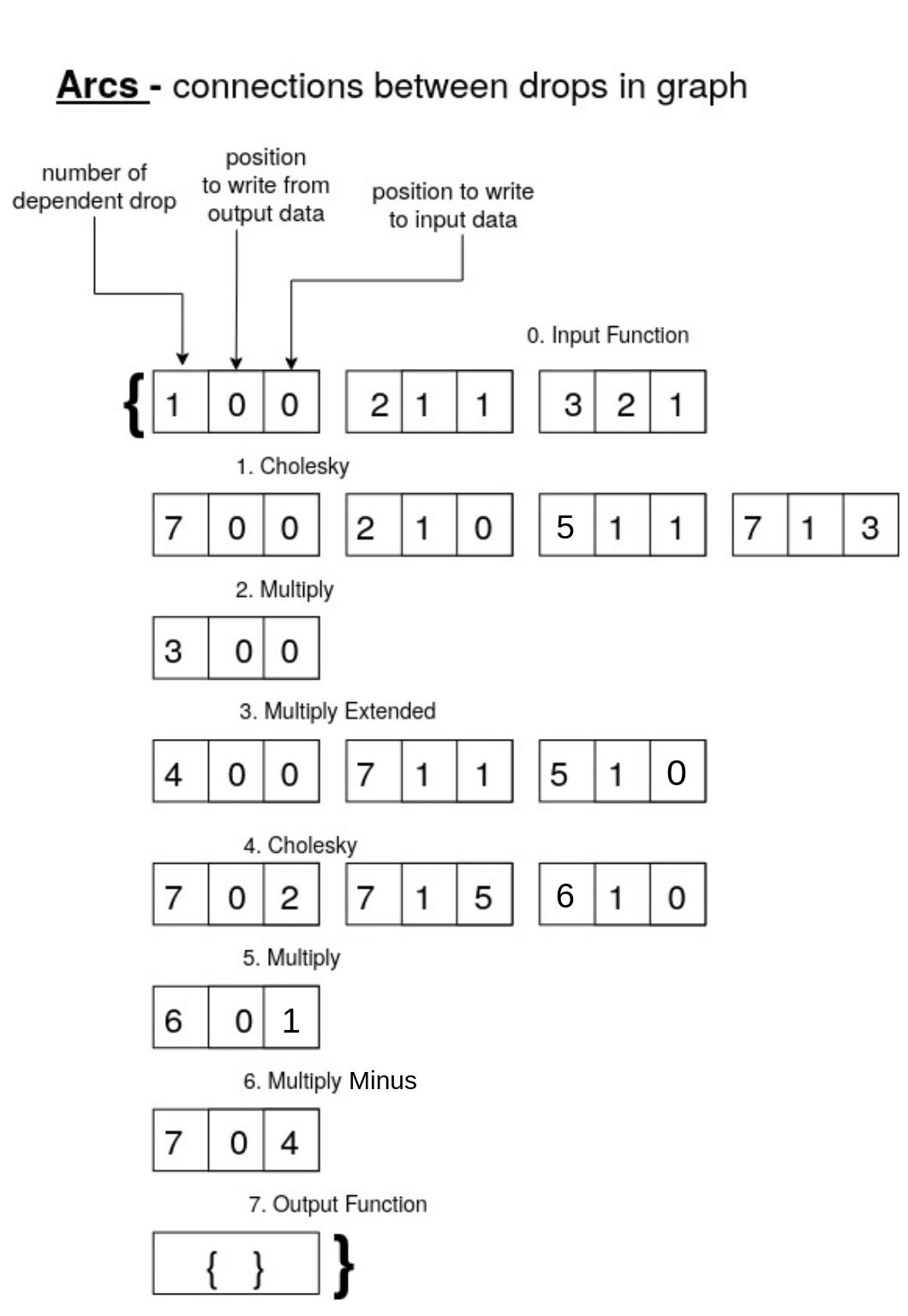}%width=100mm
 %\includegraphics[scale=0.3]{images/upload_file_button}%width=100mm
 % \caption{ Граф блочно-рекурсивного алгорит Mа  обращения треугольной  Mатрицы.}%
 \label{arcs_l}
 \end{center}
\end{figure}

Fig. 6. Description of topology of the graph of Cholesky factorization.

\subsection{Numerical Example}

Consider the Cholesky factorization of a 4 ? 4 matrix. We partition the matrix into four blocks:

$$ A =  \left(\begin{array}{cccc}16 &  24 &  28 &  4  \\ 24 &  72 &  42 &  42 \\ 28 &  42 &  85 & \ 13 \\ 4 &   42 &  13 &  74 \end{array}\right), \ 
\alpha = \left(\begin{array}{cc}16 &  24 \\ 24 &  72 \end{array}\right),   $$

$\beta = \ \left(\begin{array}{cc} 28 & \ 4 \ \\ \ 42 & \ 42 \end{array}\right), \ $
$\gamma = \ \left(\begin{array}{cc}85 & \ 13 \\ 13 & \ 74 \end{array}\right).$

1. $  (a, a_{1}) = $ \\
$$\Bigg[ \left(\begin{array}{cc}4 & \ 0 \\ 6 & \ 6 \end{array}\right) , \ \left(\begin{array}{cc}1/4 & \ 0 \\ -1/4 & \ 1/6 \end{array}\right) \Bigg] = Cholesky(\alpha)$$
 
$ \alpha_{1} = 16 ,  \beta_{1} = 24 , \gamma_{1} = 72,\ $
$$ (a', a'_{1}) = [4 , 1/4] = Cholesky(\alpha_{1}),$$
 $b' = \beta_{1}/a'=6,\ \delta_{1} = \gamma_{1} - (b')^{2}=36$, 
$$ (c', c'_{1}) = [6, 1/6] = Cholesky(\delta_{1}),$$
$z'=-c'_{1}b'a'_{1} = -1/4. $

 2. $ b^T = a_{1} \cdot \ \beta = \ \left(\begin{array}{cc}1/4 & \ 0 \\ -1/4 & \ 1/6 \end{array}\right) \ \left(\begin{array}{cc} 28 & \ 4 \ \\ \ 42 & \ 42 \end{array}\right)$
 $ = \ \left(\begin{array}{cc} 7 & \ 1 \ \\ \ 0 & \ 6 \end{array}\right) \ ; b = \ \left(\begin{array}{cc} 7 & \ 0 \ \\ \ 1 & \ 6 \end{array}\right) \ $

$ \ $

 3. $\delta = \gamma - b \cdot \ b^T = \ \left(\begin{array}{cc}85 & \ 13 \\ 13 & \ 74 \end{array}\right) \ - \ \ \left(\begin{array}{cc} 7 & \ 0 \ \\ \ 1 & \ 6 \end{array}\right) \ \left(\begin{array}{cc} 7 & \ 1 \ \\ \ 0 & \ 6 \end{array}\right) \ = \ \left(\begin{array}{cc} 36 & \ 6 \ \\ \ 6 & \ 37 \end{array}\right) \ $

 4. $(c, c_{1}) = $  
 $$\Bigg[\left(\begin{array}{cc}6 &  0 \\ 1 &  6 \end{array}\right) , \  \left(\begin{array}{cc}1/6 &  0 \\ -1/36 &  1/6 \end{array}\right)\Bigg]= Cholesky(\delta)  $$
 $ \alpha_{2} = 36 , \beta_{2} = 6 ,\gamma_{2} = 37$, 
$$ (a'', a''_{1}) =[6 , 1/6]= Cholesky(\alpha_{2}), $$
 $b'' = \beta_{2}/a''=1,\ \delta_{2} = \gamma_{1} - (b'')^{2}=36 $, 
$$  (c'', c''_{1}) = [6, 1/6]= Cholesky(\delta_{2}), $$
$ z'' = -c''_{2}b''a''_{2} =  -1/6 \cdot 1 \cdot \ 1/6 = -1/36$.

 5. $z =  - \ \left(\begin{array}{cc}1/6 & \ 0 \\ -1/36 & \ 1/6 \end{array}\right) \ \left(\begin{array}{cc} 7 & \ 0 \ \\ \ 1 & \ 6 \end{array}\right) \ $ 
 $\left(\begin{array}{cc}1/4 & \ 0 \\ -1/4 & \ 1/6 \end{array}\right) \ = \ \left(\begin{array}{cc}-7/24 & \ \ 0 \\ 37/144 & \ -1/6 \end{array}\right) \ $

$ L = \ \left(\begin{array}{cc} a &  0  \\  b &  c \end{array}\right)  =  \left(\begin{array}{cccc}4 &  0 & 0 & 0 \\ 6 &  6 &  0 & 0 \\  7 &  0  &  6 &  0 \\  1 &  6 & 1 & 6 \end{array}\right) ,\ L^{-1} =$

$$\left(\begin{array}{cc} a_{1} & 0 \\ z & c_{1} \end{array}\right) \ = \   \left(\begin{array}{cccc}\frac 1 4 & 0 &  0 &  0 \\ - \frac 1 4 & \frac 1 6 &   0 & 0  \\ -\frac 7 {24} &  0 & \frac 1 6 & 0  \\ \frac {37}{144} &  -\frac 1 6 &  -\frac 1 36 & \frac 1 6 \end{array}\right). $$

 %  \medskip

\section{EXPERIMENTS}

The computational experiments were carried out on the cluster MVS-10P CASCADE LAKE of the RAS Supercomputer Center (JSCC RAS) (2256 cores, Intel Xeon Platinum 8268, 28 cores / 48 threads, 2.9 GHz, 35.75 MB of cache memory, 96 GB RAM per processor).

\subsection{Accumulation of Computational Error in the Cholesky Algorithm}

  We carried out a series of experiments to investigate the accumulation of error in the Cholesky algorithm \cite{8}. 

It is known that in all direct methods the error increases as the matrix size grows and the total number or operations increases. For the experiments, we used triangular matrices L with integer coefficients in the interval [1, 9] and multiplied them by the transpose matrices $L^T$. Then, the matrix $A= L \cdot L^T$   was factorized. The resulting matrix $L'$ was subtracted from the original matrix to obtain the error matrix: $S = L'-L$. The greatest in absolute value element of S is the computational error in the experiment.

We used double precision floating point numbers (IEEE 754 standard) in the first series of experiments and BigDecimal numbers with 100 and 500 decimal digits in the second and third series of experiments.

Note that the package of arithmetic operations for BigDecimal numbers enables the user to specify a desired number of valid decimal digits in the fractional part of the number and round the last digit.

For each matrix size, we carried out a series of 100 experiments with random matrices. The mean and maximum errors in these 100 experiments are shown in Table 1.

The first two rows (M$_D$ and cp$_D$) contain the error obtained for double 
precision numbers. For the matrix of size 64, the mean error is 7.9, and the 
maximum error is 142. This is for the case when the exact values of the matrix elements do not exceed nine.
   
For comparison, the error accumulated when the BigDecimal format with 100 and 500 decimal digits is shown in Table 1: the common logarithm of the maximum error is shown in the rows
M$_{100}$ and  M$_{500}$, and the common logarithm of the mean error is shown in the rows  cp$_{100}$ and cp$_{500}$.

\begin{table}[H] 
%\caption{ Погрешность вычислений алгорит Mа Холецкого} % title of Table 
\centering      % used for centering table 
% { \footnotesize
\begin{tabular}{c c c c c c}  % centered columns (4 columns) 
\hline\hline                        %inserts double horizontal lines 
size & $4 $ & $8 $ & $16 $ & $32 $ & $64 $  \\  % inserts table 
%heading 
\hline                    % inserts single horizontal line 
M$_D$ & $2\cdot 10^{-13}$  & $1\cdot 10^{-10}$ &$3\cdot 10^{-6}$  & 0.6 & 142  \\    % inserting body of the table 
cp$_D$ &$6\cdot 10^{-15}$  & $4\cdot 10^{-12}$ & $6\cdot 10^{-8}$  & 0.01 & 7.9  \\
  M$_{100}$ &-96  & -93.22 & -89.3   & -79.9 & -72.1 \\
 cp$_{100}$ &-97  & -95 & -91   & -81 & -73.9 \\
  M$_{500}$ &-497 & -491 & -489   & -482 & -470  \\ 
 cp$_{500}$ &-498 & -493 & -491   & -484 & -472  \\  [1ex]  
\hline     %inserts single line 
\end{tabular} 
%   }
\label{table:nonlin}  % is used to refer this table in the text 
\end{table}

Table 1. Computational error for the Cholesky algorithm for matrices of sizes 4, 8, 16, 32, and 64

\medskip
 
   For the matrix of size 64, the common logarithm of the maximum error is  72 and 470 for numbers with 100 and 500 decimal digits, respectively.

 The experiments showed that the use of double precision for matrices of size 64 and greater is senseless due to a large computational error. For such matrices, BigDecimal must be used.
 
The detailed description of the experiments makes it easy to repeat and verify them.

\subsection{Dependence of the Machine Word Length on the Matrix Size. Complexity}

   Assuming that the computational error must not exceed one since the numbers to be found are in the interval [1, 9], we conducted a series of experiments for matrices of sizes 128, 256, 512, and 1024. The selection of the minimum number of decimal digits that are required for storing the numbers showed that it increases linearly as the matrix size grows.
   
For a matrix of size 2048,  700 decimal digits are required, and matrices of size 1024, 512, 256 , and 128 require 350, 180, 90, and 50, respectively, decimal digits.

Thus, the total complexity for dense matrices and standard multiplication algorithms grows as the fifth power of the matrix size: the complexity of the matrix algorithm grows as a cube, and the complexity of multiplying two numbers is quadratic.
If we apply a fast matrix multiplication algorithm $(O(n^\omega))$ and the 
fast multiplication of numbers $(n^{(O(1)})$, Then the total complexity can 
be reduced. For example, in the case of Strassen's matrix multiplication algorithm and Karatsuba's number multiplication  \cite{18}, \cite{19}, we obtain an algorithm with complexity $\sim n^{(\log_2 7\cdot\log_2 3)}$.  

\subsection{Main Series of Experiments. Scalability}

\subsubsection{Transmission loss ratio TLR.}

To give a numerical assessment of the scalability of the proposed solutions, we use the transmission loss ratio (TLR).
We carried out the experiments with matrices the size of which doubles. Then, the number of machine operations increases by a factor of eight. Correspondingly, the number of cores was also increased by a factor of eight for each matrix size. If the data transmission time were zero and there were no time loss due to nonuniform workload of the hardware, the computation time would be constant $k=T2/T1=1$.  

On the other hand, if the number of operations doubles and the number of cores also doubles, but the computation time doubles, then there is no scalability.
The transmission loss ratio (TLR) is defined as the coefficient of increasing the computation time $k=T_2/T_1$, where $T2$  is the computation time required when the number of operations increases by a factor of two and the number of cores also doubles.

This coefficient should be in the range $1<k<2$,  bnand scalability is characterized by its closeness to one.
In all the experiments presented in Tables 2 -- 6, the time is given in minutes, and the corresponding TLR is shown.

\subsubsection{Cholesky factorization algorithm.}

Table 2 presents the results of three series of experiments carried out on the cluster mentioned in the beginning of Section 5. In each series, the leaf block size (LS), i.e., the size of the block that was computed on one processor, was constant. In the first series it was 64, in the second one 128, and in the third series --- 256.

\begin{table}[H] 
%\caption{  } %\hbox{ Масштабируе Mость алгорит Mа Холецкого для чисел типа BigDecimal с 100 десятичны Mи знака Mи. }} % title of Table 
\centering      % used for centering table 
 { \footnotesize
\begin{tabular}{c c c c c c c c}  % centered columns (4 columns) 
\hline\hline                        %inserts double horizontal lines 
\#cores  & 1 &  & 8  &  & 64  &  & 512  \\   
% matrix  &  &  &  &  &  &  &   \\  
size  & 256 &  & 512 &  & 1024 &  & 2048  \\  
LS  &  & TLR &   & TLR &   & TLR &   \\ [0.5ex]  
%heading 
\hline                    % inserts single horizontal line 
  64 & 1.22  & 1.14 & 1.80  & 2.4 & 24.84  &1.05 &28.92\\    % inserting body of the table 
 128 &1.21  & 1.44 & 3.64  & 1.35 & 8.96  &1.31 &19.96\\ 
 256 &1.15  & 1.96 & 8.71   & 1.21 & 15.5  &1.16 &23.96\\  [1ex]       % [1ex] adds vertical space 
\hline     %inserts single line 
\end{tabular} 
}
\label{table:nonlin}  % is used to refer this table in the text 
\end{table}

Table 2. Scalability of the Cholesky algorithm for BigDecimal numbers with 100 decimal digits

\medskip

 In each series, a constant computational load per one core was maintained: when the matrix size doubled, the number of cores was increased by a factor of eight. Therefore, the number of cores was 1, 8, 64, and 512; and the matrix size was 256, 512, 1024, and 2048. BigDecimal numbers with 100 decimal digits in the fractional part were used.

  The experiments showed that the best leaf block for the cluster was 128.

  \begin{table}[ht] 
%\caption{Масштабируе Mость алгорит Mа Холецкого для разреженных  Mатриц плотности 3\%, 30\% и 100\% с числа Mи и Mеющи Mи двойную точность.} % title of Table 
\centering      % used for centering table 
\begin{tabular}{c c c c c c}  % centered columns (4 columns) 
\hline\hline                        %inserts double horizontal lines 
\#cores & 4 &  & 32  &  & 256    \\   
matrix size  & 512 &  & 1024 &  & 2048   \\ 
LS  & 32 &  & 64 &  & 128   \\  
density  &  & TLR &   & TLR &    \\ [0.5ex]  
%heading 
\hline                    % inserts single horizontal line 
3\% & 0.01  & 1.82 & 0.06  & 1.65 & 0.27  \\    % inserting body of the table 
30\% &0.019  & 1.65 & 0.085  & 1.21 & 0.15 \\ 
100\% &0.018  & 1.58 & 0.071   & 1.22 & 0.13  \\  [1ex]       % [1ex] adds vertical space 
\hline     %inserts single line 
\end{tabular} 
\label{table:nonlin}  % is used to refer this table in the text 
\end{table}

Table 3. Scalability of the Cholesky algorithm for sparse matrices with the densities 3, 30, and 100\% for double precision elements

\medskip
    
It is seen from the second row in Table 2 that, as the number of cores increases by a factor of eight, the computation time increases by a factor of $364/121=3.0$, $896/364=2.4$, and $1996/896=2.2$, respectively. Therefore, the corresponding TLR is $[\sqrt[3]{3.0}, \sqrt[3]{2.4}, \sqrt[3]{2.2}] = [1.44, 1.35, 1.31]$. 

   This data indicates good scalability of the proposed software solution.
Another series of experiments (Table 3) was carried out for sparse matrices and double precision numbers.

\subsubsection{Matrix multiplication. }

The results of experiments presented in Tables 4 and 5 were carried out for the matrix multiplication for sparse matrices. Table 4 shows the results obtained using BigDecimal numbers with 100 decimal digits, and Table 5 shows the results obtained using double precision numbers.

\begin{table}[H] 
%\caption{Масштабируе Mость алгорит Mа  Mатричного у Mножения для разреженных  Mатриц плотности 3\%, 30\% и 100\% для чисел типа BigDecimal с 100 десятичны Mи знака Mи.} % title of Table 
\centering      % used for centering table 
\begin{tabular}{c c c c c c}  % centered columns (4 columns) 
\hline\hline                        %inserts double horizontal lines 
\#cores & 4 &  & 32  &  & 256    \\   
matrix size  & 512 &  & 1024 &  & 2048   \\ 
LS  & 32 &  & 64 &  & 128   \\  
density  &  & TLR &   & TLR &    \\ [0.5ex]
%heading 
\hline                    % inserts single horizontal line 
 3\% & 0.019  & 1.85 & 0.12  & 1.7 & 0.59  \\    % inserting body of the table 
 30\% &0.26  & 1.31 & 0.59  & 1.56 & 2.26 \\ 
 100\% &2.78  & 1.20 & 4.8   & 1.51 & 16.5  \\  [1ex]       % [1ex] adds vertical space 
\hline     %inserts single line 
\end{tabular} 
\label{table:nonlin}  % is used to refer this table in the text 
\end{table}

Table 4. Scalability of the Cholesky algorithm for sparse matrices with the densities 3, 30, and 100\% for BigDecimal elements with 100 decimal digits

\begin{table}[H] 
%\caption{ Масштабируе Mость алгорит Mа  Mатричного у Mножения для разреженных  Mатриц плотности 3\%, 30\% и 100\% с числа Mи двойной точности (64 бита).} % title of Table 
\centering      % used for centering table 
\begin{tabular}{c c c c c c}  % centered columns (4 columns) 
\hline\hline                        %inserts double horizontal lines 
\#cores & 4 &  & 32  &  & 256    \\   
matrix size  & 512 &  & 1024 &  & 2048   \\ 
LS  & 32 &  & 64 &  & 128   \\  
density  &  & TLR &   & TLR &    \\ [0.5ex]
%heading 
\hline                    % inserts single horizontal line 
3\% & 0.017  & 1.86 & 0.11  & 1.7 & 0.54  \\    % inserting body of the table 
30\% &0.1  & 1.59 & 0.4  & 1.52 & 1.41 \\ 
100\% &0.73  & 1.3 & 1.6   & 1.4 & 4.36  \\  [1ex]       % [1ex] adds vertical space 
\hline     %inserts single line 
\end{tabular} 
\label{table:nonlin}  % is used to refer this table in the text 
\end{table}

Table 5. Scalability of the Cholesky algorithm for sparse matrices with the densities 3, 30, and 100\% with double precision elements (64 bits)

\subsubsection{Inversion of a triangular matrix.}

  The experiments the results of which are presented in Table 6 were conducted for the algorithm of inverting a triangular matrix. Sparse matrices with double precision elements and various degrees of sparseness were used.

\begin{table}[H] 
%\caption{Масштабируе Mость алгорит Mа обращения треугольных  Mатриц для разреженных  Mатриц плотности 3\%, 30\% и 100\% с числа Mи двойной точности.} % title of Table 
\centering
\begin{tabular}{c c c c c c}  % centered columns (4 columns) 
\hline\hline                        %inserts double horizontal lines 
\#cores & 4 &  & 32  &  & 256    \\   
matrix size  & 512 &  & 1024 &  & 2048   \\ 
LS  & 32 &  & 64 &  & 128   \\  
density  &  & TLR &   & TLR &    \\ [0.5ex]
%heading 
\hline                    % inserts single horizontal line 
3\% & 0.006  & 1.71 & 0.03  & 1.54 & 0.11  \\    % inserting body of the table 
30\% &0.012  & 1.74 & 0.063  & 1.23 & 0.188 \\ 
100\% &0.016  & 1.6 & 0.065   & 1.48 & 0.21  \\  [1ex]       % [1ex] adds vertical space 
\hline     %inserts single line 

\end{tabular} 

\label{table:nonlin}  % is used to refer this table in the text 
\end{table}

Table 6. Scalability of the algorithm of inverting triangular matrices with the densities 3, 30, and 100\% with double precision elements

\medskip

It is seen that, for the density 3\%, the TLR is lower than for the density 100\%. This is because in the case of low density, not so many processors are required since the total amount of data is 30 times less and the processors are underloaded.

\section{CONCLUSIONS}

 The runtime environment DAP is described. It is based on a dynamic scheme of controlling parallel computations on distributed memory and is designed for block recursive matrix algorithms. The basic objects, their fields, and functions are described. Operation of the two-thread system is described. The runtime environment can be used for all block recursive algorithms, and the data may be both dense or sparse.
 
The runtime environment is called DAP (drop-amine-pine). Its feature is that it sequentially unrolls functions in depth, keeping all states at any nesting level until all computations in the current computational subtree are completed. This allows any processor to freely switch from one subtask to another without waiting for the current subtask to complete.

An important feature of this runtime environment is protection against failures of some nodes in the process of computations. The parent node that sent a drop to its child node must get a result. However, it may obtain a message about the child node instead of the result. In this case, the drop task is redirected to another node. No additional changes on the other nodes are required. Only one subtree corresponding to this drop will be lost and recalculated.

The runtime environment was implemented in Java using OpenMPI. The experiments were carried out for the matrix multiplication, inversion of matrices, and Cholesky factorization algorithms. The experiments demonstrated high scalability. The maximum number of cores used in the experiments on the cluster MVS-10P was 512.

%%%%%%%%%%%%%%%%%%%%%%%%%%%%%%%%%%%%%%%%

    \end{document}